\documentclass[12pt]{article}
\usepackage{a4wide,epsfig}

\renewcommand{\thefootnote}{\fnsymbol{footnote}}
\begin{document}    

\title{\vskip-3cm{\baselineskip14pt
\centerline{\normalsize\hfill BUTP--99/17}
\centerline{\normalsize\hfill TTP99--38}
\centerline{\normalsize\hfill hep-ph/9909436}
\centerline{\normalsize\hfill September 1999}
}
\vskip.7cm
Second Order Corrections to the Muon Lifetime and the Semileptonic $B$ Decay
\vskip.3cm
}
\author{
{T. Seidensticker}$^a$
\,and
{M. Steinhauser}$^b$
  \\[3em]
  { (a) Institut f\"ur Theoretische Teilchenphysik,}\\
  { Universit\"at Karlsruhe, D-76128 Karlsruhe, Germany}
  \\[.5em]
  { (b) Institut f\"ur Theoretische Physik,}\\ 
  { Universit\"at Bern, CH-3012 Bern, Switzerland}
}
\date{}
\maketitle

\begin{abstract}
\noindent
In this article two-loop QED corrections to the muon decay and corrections of 
order $\alpha_s^2$ to the semileptonic decay of the bottom quark are 
considered. We compute the imaginary part of the four-loop diagrams 
contributing to the corresponding fermion propagator in the limit of small 
external momentum. The on-shell condition is obtained with the help of a 
conformal mapping and Pad\'e approximation. Via this method we confirm the 
existing results by an independent calculation.
\end{abstract}


\renewcommand{\thefootnote}{\arabic{footnote}}
\setcounter{footnote}{0}

\section{Introduction}
The Fermi coupling constant, $G_F$, constitutes together with 
the electromagnetic coupling constant and the mass of the $Z$ boson
the most precise input parameters of the Standard Model of elementary
particle physics. $G_F$ is defined through the muon lifetime, and the decay
of the muon, as a purely leptonic process, is rather clean --- both 
experimentally and theoretically. The one-loop corrections of order $\alpha$
were computed more than 40 years ago~\cite{KinSir59Ber58} whereas only 
recently the two-loop corrections of order $\alpha^2$ have been 
evaluated~\cite{RitStu99}. The large gap in time shows that this calculation 
is highly non-trivial. The inclusion of the two-loop terms removed the relative
theoretical error of $1.5\times 10^{-5}$ which was an estimate on the size of 
the missing corrections. The remaining error on $G_F$ now reads  
$0.9\times 10^{-5}$ and is of pure experimental nature. Upcoming experiments 
will further improve the accuracy of the muon lifetime measurement and therefore 
the  ${\cal O}(\alpha^2)$ corrections to the muon decay are very important and 
constitute a crucial ingredient from the theoretical side. These facts make it
desirable to have an independent check on the correctness of the ${\cal
O}(\alpha^2)$ result. We also want to mention that in a recent 
article~\cite{FerOssSir99} optimization methods have been used in order to 
estimate the coefficient of order $\alpha^3$.

In view of the upcoming $B$ physics experiments the evaluation of
quantum corrections to $B$ meson properties have become topical. In particular
it is possible to use the semileptonic decay rate of the bottom quark in order
to determine the Cabibbo-Kobayashi-Maskawa (CKM) matrix elements with quite 
some accuracy. In~\cite{CzaMel98} an approximate expression for the 
${\cal O}(\alpha_s^2)$ corrections of $\Gamma(b\to cl\bar{\nu}_l)$
has been obtained where a non-vanishing charm quark mass has been included.
In the decay $b\to ue\bar{\nu}_e$ the mass of the $u$ quark can be neglected
which reduces the calculation to the situation given in the muon decay.
The only additional diagrams are those which arise due to the non-abelian
structure of QCD. The results of order $\alpha_s^2$ have been obtained
in~\cite{Rit99}. The corrections proportional to the number of light quarks 
have already been computed in~\cite{LukSavWis95}. 

In this letter we confirm through an independent calculation the results of 
order $\alpha^2$ to the muon decay~\cite{RitStu99} and the one of order 
$\alpha_s^2$ to the semileptonic bottom quark decay~\cite{Rit99}. In the next 
section our method is discussed, in Section~\ref{sec:res} the results are 
presented.


\section{Method and Notation}

In~\cite{RitStu99,Rit99} the imaginary part of the muon propagator is computed
up to the four-loop level. Recurrence relations based on the 
integration-by-parts technique~\cite{CheTka81} are used in order to reduce the 
integrals to be evaluated to a minimal set --- so-called master integrals. 
They are finally evaluated by computing expansions in the ratio of external 
momentum and internal muon mass. It is possible to take the on-shell limit and
perform the infinite sum which finally leads to an exact result for the 
integrals. For concise reviews of expansion methods see e.g. \cite{Rep,Smi95}.

In contrast to~\cite{RitStu99,Rit99} our approach is based on an expansion of
the full fermion propagator in the limit
\begin{eqnarray}
  M^2 &\gg& q^2 \,,
\end{eqnarray}
where $q$ is the external momentum and $M$ is the propagator mass of the
muon and bottom quark, respectively\footnote{There are, of course, diagrams 
that do not contain an internal propagator of mass $M$. These diagrams are
computed without any expansion.}. Throughout the whole paper we will neglect 
effects induced by the non-vanishing electron and up quark mass. The on-shell 
limit $q^2 \to M^2$ will be performed afterwards with the help of Pad\'e 
approximations. This, of course, only provides an approximation to the exact
result. However, the integrals to be evaluated are simplified considerably. We
will demonstrate that the accuracy obtained with our method is sufficient to 
check the existing results and enables the same reduction of the theoretical 
error on $G_F$.

The notation is essentially adopted from~\cite{CheHarSeiSte99}, where
corrections of ${\cal O}(\alpha_s^2)$ to the decay $t \to W b$ have been
computed. For completeness we briefly repeat in this paper the main formulae.
The decay rate --- both for the muon and the bottom quark --- can be written 
in the form
\begin{eqnarray}
  \Gamma &=& 
  2 M \, \mbox{Im} \left[ z \, S_V^{OS} - S_S^{OS}
  \right]\bigg|_{z=1}, 
\label{eq:gam}
\end{eqnarray}
where
\begin{eqnarray}
  S_S^{OS} \,\,=\,\, Z_2^{OS} Z_m^{OS} \left( 1 - \Sigma_S^0 \right)
  \,,&&
  \qquad
  \label{eq:sssv}
  S_V^{OS} \,\,=\,\, Z_2^{OS}\left(1+\Sigma_V^0\right)
  \,,
\end{eqnarray}
are functions of the variable
\begin{equation}
  z = {q^2\over M^2}\,.
\end{equation}
$M$ is the on-shell mass.
$\Sigma_S^0$ and $\Sigma_V^0$ represent the scalar and vector part of the 
corresponding fermion propagator, respectively. They are functions of the 
external momentum $q$ and the bare mass $m^0$ of the fermion under
consideration. In our case they further depend on the bare electromagnetic 
coupling $\alpha^0$ and the strong coupling constant $\alpha_s^0$, 
respectively, and are proportional to the square of the Fermi coupling 
constant, $G_F^2$.

The mass renormalization constant, $Z_m^{OS}$, entering in~(\ref{eq:sssv}) can
be extracted from~\cite{GraBroGraSch90}. In contrast, $Z_2^{OS}$ has to be
evaluated in the limit $M^2\gg q^2$ since the handling of $Z_2^{OS}$ is 
determined by the computation of the fermion propagator. As we are only 
interested in the imaginary part and furthermore consider only QED or QCD 
corrections to the leading order term the quantity $Z_2^{OS}$ has to be known 
up to two loops only. The result can be taken from~\cite{CheHarSeiSte99}. 
The renormalization of $\alpha$ and $\alpha_s$ proceeds in the usual way where 
we have chosen to renormalize also the electromagnetic coupling in a first step
in the $\overline{\rm MS}$ scheme.

In order to get reliable results it is necessary to compute as many terms as
possible in the expansion parameter $z$. Subsequently a Pad\'e approximation 
is applied which is at length described in~\cite{CheHarSeiSte99}. We just want
to mention that before the Pad\'e procedure a conformal mapping can be used 
which maps the complex $z$-plane into the interiour of the unit circle. 
Following Ref.~\cite{CheHarSeiSte99} we denote those results by 
$\omega$-Pad\'es and the ones obtained without conformal mapping by 
$z$-Pad\'es.

Some Pad\'e approximants develop poles inside the unit circle ($|z|\le1$ and 
$|\omega|\le1$, respectively) in conflict with the analycity of the exact 
result. In general we will discard such numbers in the following. In some 
cases, however, the pole coincides with a zero of the numerator up to several
digits accuracy, and these Pad\'e approximations will be included in our sample.
To be precise: in addition to the Pad\'e results without any poles inside the 
unit circle, we will use the ones where the poles are accompanied by zeros 
within a circle of radius 0.01, and the distance between the pole and the 
physically relevant point $q^2/M^2=1$ is larger than 0.1.

The central values and the estimated uncertainty will be extracted from Pad\'e
results $[m/n]$ with $m+n$ not too small and $|m-n|\le 2$. The central value is 
obtained by averaging the Pad\'e results and the uncertainty is given by the 
maximal deviation from the central value.

\begin{figure}[t]
\begin{center}
\epsfig{bbllx=108,bblly=594,bburx=508,bbury=663,file=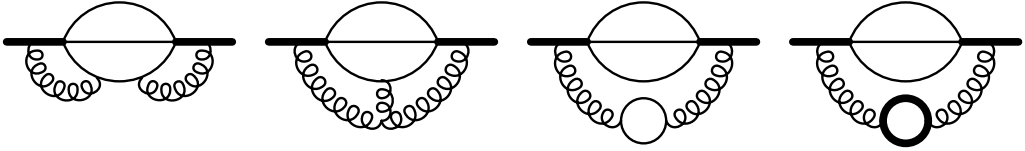}
\end{center}
\caption{\label{fig:bdec}Sample diagrams for the fermion self energy.
In the case of the bottom quark (thick line) decay two of the thin
lines represent the lepton pair and the third one the up quark. All
(one-particle-irreducible) diagrams involving the coupling of the
gluons (loopy lines) to the up and bottom quark have to be taken into
account. The modifications in the case of the muon decay are obvious.
}
\end{figure}

It is convenient to parameterize the radiative corrections for the semileptonic
bottom quark decay in the following form:
\begin{eqnarray}
  \Gamma(b\to ue\bar{\nu}_e) &=& \Gamma^0_b\left[
    A_b^{(0)}
    +\frac{\alpha_s}{\pi} C_F A_b^{(1)}
    +\left(\frac{\alpha_s}{\pi}\right)^2 A_b^{(2)}
    +\ldots
  \right]
  \,,
  \nonumber\\
  A_b^{(2)} &=&
  C_F^2 A_{b,A}^{(2)}
  +C_A C_F A_{b,NA}^{(2)}
  +C_F T n_l A_{b,l}^{(2)}
  +C_F T A_{b,F}^{(2)}
  \,,
  \label{eq:gamparb}
\end{eqnarray}
with $\Gamma^0_b=G_F^2 M_b^5 |V_{ub}|^2/(192\pi^3)$. For QCD the colour factors
are given by $C_F = 4/3$, $C_A = 3$, and $T=1/2$. $n_l$ is the number of 
massless quark flavours and will be set to $n_l=4$ at the end. $A_{b,A}^{(2)}$
corresponds to the abelian part already present in QED, $A_{b,NA}^{(2)}$ 
represents the non-abelian contribution, and $A_{b,l}^{(2)}$ and 
$A_{b,F}^{(2)}$ denote the corrections involving a second fermion loop with 
massless and heavy quarks, respectively. In principle there is also a 
contribution involving a virtual top quark loop. It is, however, suppressed by
$M_b^2/M_t^2$ and will thus be neglected. In Fig.~\ref{fig:bdec} a 
representative diagram for each one of these four colour structures is 
pictured. In Eq.~(\ref{eq:gamparb}) $\alpha_s=\alpha_s(\mu)$ is defined with 
five active flavours. The analytic result for $A_b^{(2)}$ can be found 
in~\cite{Rit99}.

An analogous expression to~(\ref{eq:gamparb}) can also be defined
in the case of the muon decay
\begin{eqnarray}
  \Gamma(\mu\to \nu_\mu e \bar{\nu}_e) &=& \Gamma^0_\mu\left[
    A_\mu^{(0)}
    +\frac{\bar\alpha}{\pi} A_\mu^{(1)}
    +\left(\frac{\bar\alpha}{\pi}\right)^2 A_\mu^{(2)}
    +\ldots
  \right]
  \,,
  \nonumber\\
  A_\mu^{(2)} &=&
  A_{\mu,\gamma\gamma}^{(2)}
  + A_{\mu,e}^{(2)}
  + A_{\mu,\mu}^{(2)}
  \,,
  \label{eq:gamparmu}
\end{eqnarray}
with $\Gamma^0_\mu=G_F^2 M_\mu^5 /(192\pi^3)$.
$\bar{\alpha}=\bar{\alpha}(\mu)$ represents the electromagnetic coupling in
the $\overline{\rm MS}$ scheme.
$A_{\mu,\gamma\gamma}^{(2)}$ represents the purely photonic corrections
whereas $A_{\mu,e}^{(2)}$ and $A_{\mu,\mu}^{(2)}$ contain an additional
electron and muon loop, respectively.
The contribution involving a virtual $\tau$ loop is not listed
in~(\ref{eq:gamparmu}) as it is suppressed by $M_\mu^2/M_\tau^2$ and
almost four orders of magnitudes smaller than the other
terms~\cite{RitStu98}.

Let us in the following describe our method used for the practical
calculation in the case of the $\mu$ decay.
The difference to
the quark decay consists only in the transition from QED to
QCD which increases the number of diagrams and makes it necessary to include
the colour factors; the idea is, however, applicable in the very
same way.

Following common practice, we investigate the effective theory where the $W$ 
boson is integrated out. The QED corrections to the resulting Fermi contact 
interaction were shown to be finite to all orders~\cite{BerSir62}. It is quite
advantageous to perform a Fierz transformation which for a pure $V-A$ theory
has the consequence that afterwards the two neutrino lines appear in the same 
fermion trace. Thus the QED corrections only affect the fermion trace involving
the muon and the electron. This also provides some simplifications in the
treatment to $\gamma_5$ since in the case of vanishing electron mass a fully 
anticommuting prescription can be used. 

As described above we consider the fermion two-point functions and compute the
imaginary part arising from the intermediate states with two neutrinos and the
electron. As a consequence already for the Born result a two-loop diagram has 
to be considered. However, it turns out that the loop integration connected to
the two neutrino lines can be performed immediately as it constitutes a 
massless two-point function. This is also the case after allowing for 
additional photonic corrections. As a result one encounters in the resulting 
diagram a propagator with one of the momenta raised to power $\varepsilon$
where $D=4-2\varepsilon$ is the space-time dimension. This slightly increases
the difficulty of the computation of the resulting diagrams. Especially for the
order $\alpha^2$ corrections, where the original four-loop diagrams are reduced 
to three-loop ones with non-integer powers of denominators, it is a priori not
clear that these integrals can be solved analytically. However, it turns out 
that for the topologies needed in our case this is indeed possible. For the 
computation of the massless two-point functions we have used the package 
{\tt MINCER}~\cite{mincer}. Slight modifications enabled us to use this package 
also for the computation of the new type of integrals.

The calculation is performed with the help of the package 
{\tt GEFICOM}~\cite{geficom}. It uses {\tt QGRAF}~\cite{Nog93} for the 
generation of the diagrams and {\tt EXP}~\cite{Sei:dipl} for the application of
the asymptotic expansion procedures. For more technical details we refer to a 
recent review concerned with the automatic computation of Feynman 
diagrams~\cite{HarSte98}.

The dispersive and absorbtive part of the fermion self energies are gauge
dependent for $q^2 \neq M^2$. Hence the dependence on the gauge parameter 
$\xi$ in Eq.~(\ref{eq:gam}) only drops out after summing infinitely many terms
in the expansion around $z=0$ and setting $z=1$. Since we are only dealing with 
a limited number of terms, our approximate results will still depend on the 
choice of $\xi$ even after taking $z \to 1$.  It is clear that with extreme 
values of $\xi$ almost any number could be produced. Thus the question arises 
which value of $\xi$ should be taken in order to arrive at a reliable 
prediction for the decay rates. As the one-loop corrections can be evaluated 
for an arbitrary gauge parameter an extensive study can be performed and we 
can gain some hints for the choice of $\xi$ at order $\alpha^2$.


\section{Results}
\label{sec:res}

Let us in a first step present the results for the muon decay
and afterwards discuss the additional diagrams necessary for the QCD
corrections to the bottom quark decay.

The lowest order (Born) diagram can be computed directly. In this case the 
electron mass can be chosen different from zero and an expansion in 
$M_e^2/M_\mu^2$ can be performed reproducing the exact result
\begin{eqnarray}
  A_\mu^{(0)} &=& 1 - 8 \, \frac{M_e^2}{M_\mu^2} 
  - 12 \, \frac{M_e^4}{M_\mu^4} \, \ln \frac{M_e^2}{M_\mu^2} 
  + 8 \, \frac{M_e^6}{M_\mu^6} - \frac{M_e^8}{M_\mu^8}
  \,.
\end{eqnarray}

We want to demonstrate the power of our method for the order $\alpha$
corrections where four three-loop diagrams contribute. At this order we 
are able to evaluate a large number of moments which gives us a suggestion 
how many terms are necessary at ${\cal O}(\alpha^2)$ in order to obtain 
reliable results. Furthermore the computation can be performed for arbitrary 
gauge parameter which also provides some hints for the two-loop QED 
corrections.

Applying the asymptotic expansion in the limit $M_\mu^2\gg q^2$ to the four
three-loop diagrams contributing to the ${\cal O}(\alpha)$ correction leads to 
the following result for the first nine expansion terms
\begin{eqnarray}
\lefteqn{A_{\mu,exp}^{(1)} = - {11 \over 8} + {25 \over 48} \, \xi
 + z_\mu \, \left(  - {61 \over 400} - {1 \over 5} \, \xi \right)
 + z_\mu^2 \, \left(  - {47 \over 540} - {1 \over 12} \, \xi \right) }
\nonumber\\&&\mbox{}
 + z_\mu^3 \, \left(  - {6929 \over 132300} - {1 \over 21} \, \xi \right)
 + z_\mu^4 \, \left(  - {11923 \over 352800} - {1 \over 32} \, \xi \right)
 + z_\mu^5 \, \left(  - {439213 \over 19051200} - {1 \over 45} \, \xi \right)
\nonumber\\&&\mbox{}
 + z_\mu^6 \, \left(  - {156487 \over 9525600} - {1 \over 60} \, \xi \right)
 + z_\mu^7 \, \left(  - {931367 \over 76839840} - {1 \over 77} \, \xi \right)
\nonumber\\&&\mbox{}
 + z_\mu^8 \, \left(  - {216409 \over 23522400} - {1 \over 96} \, \xi \right)
 + {\cal O}(z_\mu^9)
 \,.
 \label{eq:A1exp}
\end{eqnarray}

{\footnotesize
\begin{table}[t]
\begin{center}
\begin{tabular}{|l|l||c|c|c|c|c|c|c|}
\hline
input & P.A. & $\xi=-2$ & $\xi=-1$ & $\xi=-1/2$ & $\xi=0$ & $\xi=1/2$ &
$\xi=1$ & $\xi=2$ \\
\hline
7 & [4/2] & $-$1.899 & $-$1.857 & $-$1.821 & $-$1.792 & $-$1.763 & $-$1.735 & $-$1.679 \\
7 & [3/3] & $-$1.894 & $-$1.854 & $-$1.821 & $-$1.791 & $-$1.769 & $-$1.741 & $-$1.688 \\
7 & [2/4] & $-$1.900 & $-$1.858 & $-$1.822 & $-$1.792 & $-$1.765 & $-$1.736 & $-$1.673 \\
\hline
8 & [4/3] & $-$1.880 & --- & $-$1.821 & $-$1.804 & $-$1.777 & $-$1.755 & $-$1.712 \\
8 & [3/4] & $-$1.881 & --- & $-$1.821 & $-$1.805 & $-$1.777 & $-$1.753 & $-$1.711 \\
\hline
9 & [5/3] & $-$1.870 & --- & $-$1.821 & $-$1.802 & $-$1.783 & $-$1.764 & $-$1.728 \\ 
9 & [4/4] & $-$1.868 & --- & $-$1.821$^{(\star)}$ & $-$1.802 & $-$1.785 &
            $-$1.767 & $-$1.732 \\
9 & [3/5] & $-$1.871 & --- & $-$1.821 & $-$1.802 & $-$1.783 & $-$1.763 & $-$1.726 \\
\hline
\multicolumn{2}{|c||}{exact:} & \multicolumn{7}{|c|}{$-$1.810} \\
\hline
\end{tabular} 
\end{center}
\caption{\label{tab:Oal}$z$-Pad\'e results for the ${\cal O}(\alpha)$ 
  corrections for different choices of $\xi$.}
\end{table}
}

In Tab.~\ref{tab:Oal} results for the ${\cal O}(\alpha)$ coefficient can be 
found where the Pad\'e approximation is performed in the variable 
$z_\mu = q^2/M_\mu^2$. Furthermore the gauge parameter defined through the 
photon propagator $i(-g^{\mu\nu}+\xi q^\mu q^\nu/q^2)/(q^2+i\epsilon)$ is 
varied between\footnote{Despite the fact that for $\xi>1$ the generating 
functional is in principle not defined we decided to choose this range for 
the gauge parameter.} $\xi=-2$ and $\xi=+2$. Pad\'e results which develop poles
for $|z_\mu|\le1$ are in general represented by a dash. If an approximate 
cancellation with a zero from the numerator takes place (see the discussion 
above), they are marked by a star ($\star$). The comparison with the exact 
result shows that for all values of $\xi$ reasonable agreement is found. 
However, there is a clear preference for $\xi=0$ where the agreement with the 
exact result is best\footnote{It seems that the $\xi$-dependent terms of 
$A_{\mu,exp}^{(1)}$ follow from the construction rule 
$25/48 - \sum_{n=1}^\infty z_\mu/n/(n+4)$ which for $z_\mu = 1$ indeed gives 
zero.}. Thus we will adopt this value for the ${\cal O}(\alpha^2)$ calculation. 

{\footnotesize
\begin{table}[t]
\begin{center}
\begin{tabular}{|l|l||c|c|}
\hline
input & P.A. & $z$ & $\omega$ \\
\hline
7 & [4/2] & $-$1.792 & $-$1.790$^{(\star)}$ \\
7 & [3/3] & $-$1.791 & $-$1.811 \\
7 & [2/4] & $-$1.792 & $-$1.816 \\
\hline
8 & [4/3] & $-$1.804 & $-$1.803 \\
8 & [3/4] & $-$1.805 & $-$1.788 \\
\hline
9 & [5/3] & $-$1.802 & $-$1.807 \\ 
9 & [4/4] & $-$1.802 &  --- \\
9 & [3/5] & $-$1.802 & $-$1.808 \\
\hline
\multicolumn{2}{|c||}{exact:} & \multicolumn{2}{|c|}{$-$1.810} \\
\hline
\end{tabular} 
\end{center}
\caption{\label{tab:Oal_2}Pad\'e results for the ${\cal O}(\alpha)$
  corrections ($z$- and $\omega$-Pad\'es). }
\end{table}
}

In Tab.~\ref{tab:Oal_2} the gauge parameter is fixed to $\xi=0$ and in addition
to the $z$-Pad\'es also the $\omega$-Pad\'es are listed. With the inclusion
of more moments the approximation to the exact result improves. Taking only 
those results into account where eight or nine input terms enter the following
result for the order $\alpha$ correction can be deduced 
\begin{eqnarray}
  A^{(1)} &=& -1.80(1)
  \,.
  \label{eq:a1full}
\end{eqnarray}
Here the notation $1.80(1) = 1.80\pm0.01$ has been adopted. The excellent 
agreement with the exact result quoted in Tab.~\ref{tab:Oal_2} encourages the 
use of our method.

At order $\alpha^2$ only the first eight moments are at hand. Using only seven 
and eight input terms changes the numbers of Eq.~(\ref{eq:a1full}) to 
\begin{eqnarray}
  A^{(1)} &=& -1.80(2)
  \,,
\end{eqnarray}
which is still sufficiently accurate.

Let us now move on to the two-loop QED corrections. Altogether 44 four-loop 
diagrams contribute. The application of the asymptotic expansion in the limit 
$M_\mu^2\gg q^2$ leads to 72 sub- and cosub-diagrams, which have to be 
evaluated. The analytical expressions obtained from the asymptotic expansion 
are quite lengthy. Thus we refrain from listing the results explicitly and 
present them only in numerical form.

{\footnotesize
\begin{table}[t]
\begin{center}
\begin{tabular}{|l|l||c|c||c|c||c|c|}
\hline
\multicolumn{2}{|c||}{} & \multicolumn{2}{|c||}{$\gamma\gamma$} & 
  \multicolumn{2}{|c||}{$e$} & \multicolumn{2}{|c|}{$\mu$} \\
\hline
input & P.A. & $z$ & $\omega$ & $z$ & $\omega$ & $z$ & $\omega$ \\
\hline
6 & [3/2] & 3.028 & 3.876 & 2.882 & 3.237 & --- & --- \\
6 & [2/3] & 3.029 & 3.862 & 2.893 & 3.397 & $-$0.0365$^{(\star)}$ & --- \\
\hline
7 & [4/2] & 3.067 & 3.815 & 2.931 & 3.898 & $-$0.0364 & --- \\
7 & [3/3] & 3.062 & ---   & 2.969 & 3.332 & $-$0.0364 & --- \\
7 & [2/4] & 3.066 & 3.841 & 2.959 & 3.353 & --- & --- \\
\hline
8 & [4/3] & --- & 3.919$^{(\star)}$ & 3.000 & 3.287$^{(\star)}$ & 
            $-$0.0364 & --- \\
8 & [3/4] & --- & 3.869$^{(\star)}$ & 3.002 & 3.581$^{(\star)}$ & 
            $-$0.0364 & --- \\
\hline
\multicolumn{2}{|c||}{exact:} & \multicolumn{2}{|c||}{3.56} & 
 \multicolumn{2}{|c||}{3.22} & \multicolumn{2}{|c|}{$-$0.0364} \\
\hline
\end{tabular} 
\end{center}
\caption{\label{tab:Oal2}Pad\'e results for
$A_{\mu,\gamma\gamma}^{(2)}$, $A_{\mu,e}^{(2)}$ and $A_{\mu,\mu}^{(2)}$. }
\end{table}
}

Our results at ${\cal O}(\alpha^2)$ are summarized in Tab.~\ref{tab:Oal2} 
where the scale $\mu^2=M_\mu^2$ has been 
adopted\footnote{The $\ln\mu^2/M_\mu^2$ terms can be reconstructed with the 
help of the $\beta$ function governing the running of $\bar\alpha(\mu)$.}. 
For comparison in the last line the numbers presented in Ref.~\cite{RitStu99}
are listed. The results one obtains using the Pad\'e approximants computed with 
seven and eight input terms read
\begin{eqnarray}
  A_{\mu,\gamma\gamma}^{(2)} &=& 3.5(4)
   \,,
  \nonumber\\
  A_{\mu,e}^{(2)} &=& 3.2(6)
  \,,
  \nonumber\\
  A_{\mu,\mu}^{(2)} &=& -0.0364(1) 
  \,.
  \label{eq:resmu2}
\end{eqnarray}
It is remarkable that the central values agree well with the exact results 
which can be interpreted as a sign that the presented error estimations are 
quite conservative. Furthermore it can be claimed that via our method we were
able to confirm the results of~\cite{RitStu99}.

The error of $A_{\mu,\mu}^{(2)}$ is particularly small, as the expansion in 
$z$ converges very quickly. Note that in this case all $\omega$-Pad\'es develop
poles inside the unit circle. The $z$-Pad\'es are, however, very stable. A 
similar behaviour has been found for the analogous contribution to the decay
of the top quark into a $W$ boson and a bottom quark~\cite{CheHarSeiSte99}.  

{\footnotesize
\begin{table}[t]
\begin{center}
\begin{tabular}{|l|l||c|c|}
\hline
input & P.A. & $z$ & $\omega$ \\
\hline
6 & [3/2] & 5.836 & 7.249$^{(\star)}$ \\
6 & [2/3] & 5.836 & 7.057 \\
\hline
7 & [4/2] & 5.935 & 7.040 \\
7 & [3/3] & 5.833$^{(\star)}$ & 7.076 \\
7 & [2/4] & 5.938 & 7.080 \\
\hline
8 & [4/3] & 6.110 & 6.873$^{(\star)}$ \\
8 & [3/4] & 6.113  & 7.060$^{(\star)}$ \\
\hline
\multicolumn{2}{|c||}{exact:} & \multicolumn{2}{|c|}{6.743} \\
\hline
\end{tabular} 
\end{center}
\caption{\label{tab:Oal2_2}Pad\'e results for the corrections of ${\cal
    O}(\alpha^2)$ to the muon decay, $A^{(2)}_\mu$.}
\end{table}
}

A prediction for the decay rate of the muon up to order $\alpha^2$ can be in 
principle obtained by summing the individual terms of~(\ref{eq:resmu2}). This
would, however, significantly overestimate the error. It is more promising to
add in a first step the moments of the single contributions and perform 
the Pad\'e procedure for the sum. The corresponding results are shown in 
Tab.~\ref{tab:Oal2_2} which finally lead to
\begin{eqnarray}
  A_\mu^{(2)}  &=& 6.5(7) \,.
\end{eqnarray}
The deviation of the central value from the exact result of 6.743 is less
than 3\% and well covered by the extracted error of roughly 10\%. Thus the sole
knowledge of our results would also reduce the theoretical error on $G_F$ as 
mentioned in the Introduction. Using the results presented in this paper the
decay rate of the muon reads
\begin{eqnarray}
  \Gamma(\mu\to \nu_\mu e \bar{\nu}_e) &=& \Gamma^0_\mu\left[
    0.9998
    - 1.810 \frac{\bar\alpha(M_\mu)}{\pi}  
    + 6.5(7) \left(\frac{\bar\alpha(M_\mu)}{\pi}\right)^2 
    +\ldots
  \right]
  \,.
\end{eqnarray} 

As already noted in~\cite{FerOssSir99} the numerical coefficient in front of 
the second order corrections becomes very small if one uses the on-shell 
scheme for the definition of the coupling constant $\alpha$. Then the
$\overline{\mbox{MS}}$ coupling is given by 
$\bar\alpha(M_\mu)=\alpha(1+\alpha/3\pi \ln(M_\mu^2/M_e^2))$ and there is an 
accidental cancellation between the constant and the logarithm in the second
order corrections. 

Let us now turn to the semileptonic decay of the bottom quark. The Born and 
one-loop corrections can, of course, be taken from the muon decay. In 
particular we have $A_\mu^{(0)}=A_b^{(0)}$ and $A_\mu^{(1)}=A_b^{(1)}$. As far
as the two-loop terms are concerned only the non-abelian contribution,
$A_{b,NA}^{(2)}$, has to be computed in addition. The other colour structures
are related to the expressions occurring in the muon decay rate through
$A_{b,A}^{(2)}=A_{\mu,\gamma\gamma}^{(2)}$, $A_{b,l}^{(2)}=A_{\mu,e}^{(2)}$, 
and $A_{b,F}^{(2)}=A_{\mu,\mu}^{(2)}$, with obvious replacements of the masses.

{\footnotesize
\begin{table}[t]
\begin{center}
\begin{tabular}{|l|l||c|c|}
\hline
input & P.A. & $z$ & $\omega$ \\
\hline
6 & [3/2] & $-$8.374 & $-$9.253 \\
6 & [2/3] & $-$8.376 & $-$9.421 \\
\hline
7 & [4/2] & $-$8.469 & $-$9.164 \\
7 & [3/3] & $-$8.560 & $-$9.076 \\
7 & [2/4] & $-$8.476 & $-$9.288 \\
\hline
8 & [4/3] & $-$8.616 & $-$9.073 \\
8 & [3/4] & $-$8.616 & $-$9.073 \\
\hline
\multicolumn{2}{|c||}{exact:} & \multicolumn{2}{|c|}{$-$9.046} \\
\hline
\end{tabular} 
\end{center}
\caption{\label{tab:NA}Pad\'e results for the non-abelian part of ${\cal
    O}(\alpha_s^2)$, $A^{(2)}_{b,NA}$.} 
\end{table}
}

In Tab.~\ref{tab:NA} the results for $A_{b,NA}^{(2)}$ can be found.
We infer
\begin{eqnarray}
  A_{b,NA}^{(2)} &=& -8.8(4) \,,
\end{eqnarray}
where $\mu^2=M_b^2$ has been chosen. The central value is again in very good 
agreement with the exact result~\cite{Rit99} and agrees within the error
estimate of 5\%.

{\footnotesize
\begin{table}[t]
\begin{center}
\begin{tabular}{|l|l||c|c|}
\hline
input & P.A. & $z$ & $\omega$ \\
\hline
6 & [3/2] & $-$20.587 & $-$21.835 \\
6 & [2/3] & $-$20.592 & $-$22.252 \\
\hline
7 & [4/2] & $-$20.744 & $-$21.159 \\
7 & [3/3] & $-$20.836 & --- \\
7 & [2/4] & $-$20.757 & $-$21.649 \\
\hline
8 & [4/3] & $-$20.964 & $-$21.213 \\
8 & [3/4] & $-$20.965 & $-$21.608 \\
\hline
\multicolumn{2}{|c||}{exact:} & \multicolumn{2}{|c|}{$-$21.296} \\
\hline
\end{tabular} 
\end{center}
\caption{\label{tab:NA_2}Pad\'e results for the corrections of 
${\cal O}(\alpha_s^2)$ to $b \to ue\nu_e$, $A^{(2)}_b$.}
\end{table}
}

In order to get predictions for $A_b^{(2)}$ we again add the moments in a 
first step and perform the Pad\'e approximations afterwards. From the results 
listed in Tab.~\ref{tab:NA_2} we deduce
\begin{eqnarray}
  A_b^{(2)} &=&  -21.1(6)
  \,.
\end{eqnarray}
This number is in good agreement with the one stated in~\cite{Rit99}. The error
is quite small and amounts only to roughly 3\%. The semileptonic decay rate
of the bottom quark finally reads
\begin{eqnarray}
  \Gamma(b\to ue\bar{\nu}_e) &=& \Gamma^0_b\left[
    1
    - 2.413 \frac{\alpha_s(M_b)}{\pi} 
    - 21.1(6) \left(\frac{\alpha_s(M_b)}{\pi}\right)^2 
    +\ldots
  \right]
  \,.
\end{eqnarray}


\section{Conclusions}
\label{sec:con}

In this paper the two-loop QED corrections to the decay rate of the muon
have been evaluated. A new method has been used in order to confirm via an 
independent calculation the result of Ref.~\cite{RitStu99}. From the muon decay
the Fermi coupling constant, $G_F$, is determined which constitutes one of the 
basic input parameters of the Standard Model. Thus it is very important to have
independent checks on such highly non-trivial computations. The inclusion of 
the new correction terms removes the theoretical error on $G_F$. After the 
additional computation of the non-abelian diagrams the QCD corrections to the
semileptonic decay rate of the bottom quark $\Gamma(b\to ue\nu_e)$ are obtained.
Again agreement with the literature~\cite{Rit99} is found. 


\section*{Acknowledgments}
We would like to thank K.G. Chetyrkin and J.H.~K\"uhn for valuable 
discussions and careful reading of the manuscript. Useful discussions with 
T.~van~Ritbergen are gratefully acknowledged. This work was supported by the 
{\it Graduiertenkolleg ``Elementarteilchenphysik an Beschleunigern''}, the 
{\it DFG-Forschergruppe ``Quantenfeldtheorie, Computeralgebra und 
Monte-Carlo-Simulationen''} under Contract Ku 502/8-1 and the {\it Schweizer
Nationalfonds}.


\end{document}